\documentclass[aps,prc,twocolumn,showpacs,superscriptaddress,floatfix]{revtex4}

\usepackage{graphicx}
\usepackage{amsmath,amssymb}
\usepackage{times,txfonts}

\begin{document}

\title{Reply to the Comment on 'Pygmy dipole response of proton-rich argon nuclei in random-phase approximation and no-core shell model'.}

\author{C. Barbieri}
\affiliation{Gesellschaft f\"ur
  Schwerionenforschung Darmstadt, Planckstr. 1, D-64259 Darmstadt,
  Germany}
\affiliation{Theoretical Nuclear Physics Laboratory, RIKEN
  Nishina Center, Wako 351-0198, Japan}

\author{E. Caurier}
\affiliation{Institut de Recherches Subatomiques, Universit\'e Louis
  Pasteur, F-67037 Strasbourg, France}

\author{K. Langanke}
\affiliation{Gesellschaft f\"ur
  Schwerionenforschung Darmstadt, Planckstr. 1, D-64259 Darmstadt,
  Germany}
\affiliation{Institut f\"ur Kernphysik, Technische
  Universit\"at Darmstadt, Schlossgartenstr. 9, D-64289 Darmstadt,
  Germany}

\author{G. Mart\'inez-Pinedo}
\affiliation{Gesellschaft f\"ur
  Schwerionenforschung Darmstadt, Planckstr. 1, D-64259 Darmstadt,
  Germany}

\date{\today}


\pacs{21.10.Gv, 24.30.Gd, 24.10.Cn, 21.60.Cs, 21.60.Ev, }
\maketitle


In Ref.~\cite{Bar.08} we have studied the distribution of 
low-lying dipole strength in proton-rich nuclei. The focus of our work
was on possible differences in the description of this dipole response
between the random phase 
approximation (RPA), adopted in previous studies, and the 
shell model. The $^{32}$Ar and $^{34}$Ar isotopes were studied, 
in view of upcoming experimental data. The main results of our study were
that the inclusion of configuration mixing, as described by the
shell model, broadens the low-lying strength and reduces its strength.

In a comment to our work~\cite{Paa.08} Paar argues that our results
for $^{32,34}$Ar are inconsistent with a previous study~\cite{Paa.06}
that used the same $V_{\text{UCOM}}$ interaction but applied to medium
and heavy nuclei near the valley of stability. Reference~\cite{Paa.06} has
shown that the $V_{\text{UCOM}}$ is in disagreement with data on the dipole
strength distribution for {\em stable} nuclei.  In particular, the
Hartree-Fock (HF) gaps at the Fermi surface are too
large~\cite{Paa.06,Bar.06}, which translates into an overestimation of
the centroid energy of giant dipole resonances (GDR)~\cite{Pap.07a}.
Our calculations for stable nuclei confirm these results.  However,
when applying $V_{\text{UCOM}}$ to calculate the dipole strength for nuclei
close to the dripline, where the relevant degrees of freedom are
dominated by the proximity to the continuum, we find a quite different
behavior.  
Valence nucleons near the driplines have significantly smaller separation
energies which are correctly reproduced at the HF level (2.13~MeV in our
calculation for $^{32}$Ar, compared to the experimental value of 2.40~MeV).
 The next orbits are found in the continuum. Their position is dominated
by kinetic energy and results in much smaller gaps between the two major shells,
as compared to stable nuclei.
 This allows for the generation of enhanced dipole strength at low energies
and improves the description of the GDR centroid at the RPA level.

Our calculations show that applying the V$_{\text{UCOM}}$ interaction to
proton-rich Ar isotopes leads to an RPA description of the dipole
response which is: (i) in qualitative agreement with the systematic of
dripline nuclei (a low-energy peak and a giant resonance), and (ii) in
good quantitative agreement with other theoretical predictions for the
same nuclei~\cite{Paa.05} (no experimental data are yet available).
The same interaction was then applied in our shell model calculations
and a reduction of the low-energy dipole strength with respect to the
RPA study was found. This is the main result of our work published in
Ref.~\cite{Bar.08}.

As noted in the conclusions of Ref.~\cite{Bar.08}, second RPA (SRPA) 
calculations with V$_{\text{UCOM}}$ tend to noticeably lower the GDR
centroid energy,  
bringing them into better agreement with data~\cite{Pap.08}. 
This is of course the experience with stable nuclei. 
SRPA calculations of dipole response for dripline nuclei have yet not 
been performed and hence the question, whether the proximity of the continuum
might affect these calculations, is still open. However, our 
shell model study has shown that the inclusion of 2-particle-2-hole 
configurations do not alter the energies of the enhanced low-lying
strength or of the GDR centroid energy.

The term ``Pygmy'' is often used in the literature to indicate the presence 
of enhanced dipole strength at low excitation energies.  
In this sense we have used this term. 
The Comment~\cite{Paa.08} criticized our use of the notion ``pygmy resonance''
suggesting that this term is reserved for states of a clear collective
nature. As only the transition strength and transition densities could be
acceesed experimentally, any other quantity is model dependent.
 This situation has been already found in two studies of low-lying strength
for $^{132}$Sn that conclude that is either collective~\cite{Vret.01} or
non-collective~\cite{Sar.04}.

In summary, the aim of Ref.~\cite{Bar.08} was to study the influence
of correlations beyond the simple RPA approach on the low-lying dipole
response for proton-rich nuclei.  For the particular case of the
isotopes $^{32,34}$Ar, we found that the V$_{UCOM}$ interaction gives
a reasonable RPA dipole response.  As this interaction is readily
accessible for RPA and shell model calculations, we felt justified to
adopt it for the first shell model calculation of low-energy dipole
response in nuclei at the proton dripline investigating the importance
of correlations beyond RPA on the dipole response.

\end{document}